\documentclass[runningheads]{llncs}
\usepackage[T1]{fontenc}
\usepackage{amsmath,amssymb,amsfonts}
\usepackage{algorithm}
\usepackage{algpseudocode}
\usepackage{graphicx}
\usepackage{adjustbox}
\usepackage{listings}
\usepackage{eiffel}

\usepackage{hyperref}
\usepackage{color}

\usepackage{tikz}
\usetikzlibrary{positioning}
\usetikzlibrary{arrows.meta}
\tikzset{%
  >={Latex[width=3pt,length=3pt]},
            code/.style = {rectangle, rounded corners, draw=black,
                           minimum width=3pt, minimum height=3pt,
                           text centered},
            description/.style = {minimum width=3pt, minimum height=3pt,
                           text centered}
}

\usepackage{textcomp}
\usepackage{xcolor}

\newcommand{\secref}[1]{Section~\ref{#1}}
\newcommand{\figref}[1]{Figure~\ref{#1}}

\begin{document}


\title{Large Language Models and Language Server Protocol: a match made in context}
\titlerunning{LLMs and LSP: a match made in context}
\author{
Alessandro Schena\inst{1}\orcidID{0009-0005-1793-9107} \and
Ilgiz Mustafin\inst{1}\orcidID{0009-0007-0476-5966} \and
Julia Kotovich\inst{1,2}\orcidID{0000-0003-3739-6304}
}

\institute{
  Constructor Institute of Technology, Schaffhausen, Switzerland
  \email{\{Alessandro.Schena, Ilgiz.Mustafin, Julia.Kotovich\}@constructor.org}\and
  Constructor University, Bremen, Germany
}

\maketitle

\newcommand{\toolname}{Eiffel-tools}

\begin{abstract}
  This article introduces \toolname{}, a language server protocol (LSP) implementation for the Eiffel programming language that uses Large Language Models (LLMs) to aid the development of statically verified software.
  The tool provides various interactive and non-interactive commands to produce code and specifications.   It uses language and project specific knowledge to precisely direct the LLM and verifies the output using a static verifier. It crafts rich programmatic prompts for the input and corrects or rejects the output. Furthermore, it handles the retries until the program passes verification.

  The tool's bug fixing capability is evaluated on 2 public datasets using 3 models. The tool can fix 76\% to 95\% of bugs by combining LLMs and a formal verifier depending on the model and prompts used.
  The results show the trade-off between the number of fixing attempts and
  the success rate.

\end{abstract}

\section{Introduction}
Specification-driven development (SDD) is getting more attention with the growing
use of Large Language Models (LLMs) for development \cite{10.1007/978-3-031-75434-0_9}. In this approach,
the development is done in a natural language by writing specifications while
the program code is then generated automatically
from the specifications.

Expressing specifications using a formal language, even partially \cite{meyer2022handbook},
opens the door for various analysis tools. For example,
using Design by Contract \cite{design-by-contract} in Eiffel,
the program's state is checked against specifications during execution.
This allows test runs of the program to show the presence of implementation
errors.

A more advanced analysis such as formal verification allows finding inconsistencies between specifications and implementation statically, without running the code. This allows finding errors earlier and mathematically proving arbitrary functional properties. Still, errors in specifications can lead to invalid but correct programs.

Modern SDD relies heavily on LLMs. While LLMs code generation might work well for programming languages that are well-represented in the training datasets, their ecosystem might not have formal verification tools available. 
These are called high-resource languages~\cite{joel2025surveyllmbasedcodegeneration}.

Eiffel is a low-resource language; it has a first-class support for specifications and a static verifier AutoProof, but the low amount of public code makes LLM code generation more challenging. Adapter tools could mitigate this issue. Any such tool must structure and contextualize the code for the model and must keep the user in the loop whenever a reasonable output is produced.
To mitigate this issue, special tools can be implemented which take the role
of an adapter: on the one hand, the tool must structure and contextualize the
code for the model, on the other hand, the tool must present model's
outputs to allow the user to benefit more from the LLM.

This paper introduces \toolname{} \cite{eiffel-tools}, as the aforementioned adapter.
\toolname{} implements the Language Server Protocol (LSP) to integrate conveniently in the majority of code editors.

In this paper we highlight the tool's ``fix feature body'' command. 
The command tries to change the body of the procedure such that it will satisfy its specifications.
The tool presents the relevant parts of the code, while omitting irrelevant parts of the project,
to the LLM and sets up
the interaction loop between the LLM and the verifier.

We present the evaluation of this command using two datasets of buggy
programs with correct specifications and compare the performance of three LLMs.
One model is a small generalist model, second is a small coding model and another
is a bigger coding model.

The remainder of this paper is structured as follows. \secref{background} provides the necessary background, introducing the core concepts related to the Eiffel programming language, large language models, and LSP-based tooling in greater detail.
\secref{related-work} reviews related work in automated program repair, LLM-assisted development, and contract-driven methodologies.
\secref{design-and-implementation} presents the design and implementation of our experimental setup, describing how contract-guided repair is operationalized using LLMs.
\secref{evaluation} outlines the evaluation methodology, followed by \secref{results-and-discussion}, which reports the experiment results and discusses them.
\secref{limitations} addresses threats to validity and limitations of the study.
Finally, \secref{future-work} outlines directions for future work.





\section{Background}\label{background}
This section provides the background information about the Eiffel programming
language and the related technologies.

\subsection{Eiffel and AutoProof}\label{eiffel-and-autoproof}

Eiffel~\cite{eiffel_language_and_environment} is an object-oriented programming language with first-class support for Design by Contract~\cite{design-by-contract}, allowing developers to embed preconditions, postconditions, and class invariants directly in the source code. Thus, a typical Eiffel project contains both the executable program and its formal specifications within the same codebase. For this reason, we consider Eiffel to be natively suitable for SDD.

Eiffel's contracts are used commonly for dynamic checking. When a program is executed in testing or development environment, its contracts transformed in equivalent runtime assertions. For example, when there is a procedure call, the callee's preconditions are asserted in the caller just before the call. These checks are removed in the finalized version of the program.

Another use of contracts is static verification via AutoProof \cite{autoproof}. AutoProof is an auto-active verifier for the Eiffel programming language that can prove functional correctness of Eiffel programs annotated with contracts. It works only with model-based contracts. Model-based contracts refer to data structures which are formalized by axioms in the lower-level verifier. They are not executable. They are the bridge that connects Eiffel to Boogie~\cite{boogie-lang}, the underlying tool that generates proof obligations for some SMT solver. This allows proving correctness of the program for all possible execution paths, in contrast with testing where it might be impossible to test the program in all possible states.

\subsection{Large language models for software development}\label{llm}

The recent advancements in deep learning brought large language models into industrial use for software development.
LLMs need proper context to provide useful information because despite their impressively coherent outputs, they are not oracles and cannot infer user intentions that are not explicitly stated.
This paper shows that language servers are a natural place to condense language and project specific information to guide the generation of large language models.
We expect the constructed context to compensate partially the scarcity of training data for low resources languages and boost the quality of performance for all other languages.

\subsection{Language server protocol}\label{lsp}

The language server protocol (LSP)~\cite{lsp} is proposed to be used as a common integration point
between the code editor, the verifier and the LLM.
It defines a compatibility layer between source code editors and language specific functionalities.
To exemplify the widespread adoption of this protocol, let us note that at the time of this writing there are 302 server side and 65 client side implementations listed on the official website~\cite{lsp}.

When a programming editor implements the client side of the LSP, it automatically provides ``smart editing features'' for any programming language supported by a protocol compliant server.
Some of the language-specific features which the LSP standardizes are code completion, marking of lints, warnings and errors and editing routines.
The latter can be refactoring routines, formatting routines, and custom operations with domain specific knowledge.
Within the protocol they are shown as code actions or commands; the former being parametrized only by the user's cursor position and the latter which can be called from anywhere with the right arguments.

These custom operations may access the server state and use the project and language specific knowledge to implement their custom logic.
This is a perfect environment for defining guidelines for LLMs: which information is useful in the prompt, which outputs satisfy the syntax of the language and which identifiers are valid in the place the custom command wants to write and more.
In our case, because Eiffel can define formal specifications and AutoProof can prove the correctness of the code against those specifications, we can force any command that generates code to produce only verified code.

\section{Related work}\label{related-work}

The research area applying LLMs to software engineering is flourishing, with a rapidly expanding body of work addressing a wide range of development and analysis tasks. Authors of~\cite{braberman2025generativetransformationspatternsllmnative} provide a systematic taxonomy of generative transformations and their compositional patterns in LLM-native applications, primarily as an analytical framework. Another study~\cite{hou2024large} conducts a systematic literature review of LLM applications in SE, with a particular focus on understanding how LLMs can be exploited to optimize processes and outcomes. Complementary to these studies, the authors~\cite{zhang2024unifyingperspectivesnlpsoftware} systematically review recent advances in LLM applications for software engineering. They analyze applications in requirements, testing, deployment, and operations, and identify key challenges and future directions.

The study~\cite{fan2023largelanguagemodelssoftware} emphasizes on the investigation of the emerging area of LLMs for SE and their applications, highlighting the importance of hybrid techniques, such as traditional SE plus LLMs for the reliable and efficient software development. 

LSP can be used to provide context during the LLM inference through monitors~\cite{monitors}. This approach uses language servers externally, primarily leveraging their completion engines, to dynamically mask the streaming output of large language models, which further enhances fine-tuning and dynamic masking. With the same philosophy, our work defines where and how the LLMs may find language and project specific knowledge, which adds benefit to fine-tuning and dynamic masking.

CITADEL~\cite{citadel} and AutoInfer~\cite{autoinfer} automatically generate Eiffel specifications. Citadel is an Eiffel front end for Daikon~\cite{daikon}. AutoInfer~\cite{autoinfer} is another tool that generates Eiffel contracts using data mining and contract templates.
However, these tools generate specifications for dynamic checks, not model-based specifications for static verification.

On the side of automatic repair of Eiffel programs there is AutoFix~\cite{autofix} and Proof2Fix~\cite{proof2fix}. AutoFix uses random testing through AutoTest~\cite{autotest} to identify the faults. Proof2Fix identifies faults using AutoProof. They both propose candidate fixes that may be the substitution of a template for code edits or the strengthening of the preconditions or weakening of the postconditions for contracts edits. However, these tools are limited to Eiffel and do not generalize to LLM-guided, project- and language-specific specification and verification workflows, which is the focus of our approach.

An experiment similar to the evaluation of this paper is presented in \cite{le2025can}
with the focus
on generating specifications and fixing them. Similar prompting techniques are
discussed for JML and Frama-C.

A close work in the field of LLM-assisted development and manual program repair is
the study~\cite{huang2025aimodelshelpproduce} of developers producing verified fixes with and without
the assistance of an LLM chat.

\section{Design and implementation}\label{design-and-implementation}

\toolname{} is an LSP language server which provides several novel
commands using LLMs and a static verifier to generate or repair Eiffel code and
specifications.

The language server avoids using a full-fledged Eiffel compiler to be performant and provide support to projects in inconsistent states with compilation errors and missing files.
\toolname{} reads the project configuration file (ECF),
parses it, extracts the used libraries and recursively analyses them.
All classes used in the system are indexed (features, contracts, inheritance tree is stored)
and available through the usual
LSP interfaces which allows code editors, for example, to display and navigate
between classes and features in the source code. This information is also
integrated into LLM prompts when relevant to give the relevant context to the model.

Eiffel source code in files and LLM responses is parsed using an external
parser \cite{tree-sitter-eiffel} based on Tree-sitter \cite{treesitter}.
This allows using Tree-sitter queries to extract the relevant structures
from the source code. For example, when LLM is used to generate code,
the parser is used to find valid features in the responses and
then to extract code or contracts of the found feature.

The tool does not depend heavily on any specific LLM or an API.
Currently an OpenAI-like \texttt{chat/completions} API is supported.
We are using the OpenRouter platform which provides access
to several LLMs using the same API to conduct the experiment.

\toolname{} uses AutoProof to verify existing code or LLM-generated suggestions.
Because AutoProof may not terminate on some inputs, \toolname{} runs it with timeouts: any verification exceeding a time limit is killed, and if the code was LLM-generated, the tool reverts it to its previous stable state upon failure.

For organizational clarity, \toolname{} commands are classified along two dimensions:
\begin{itemize}
  \item By interaction mode
        \begin{itemize}
          \item Interactive, invoked as code actions by the LSP client
          \item Non-interactive, invoked as standalone executables
        \end{itemize}
  \item By focus
        \begin{itemize}
          \item specifications
          \item code
          \item specification and code
        \end{itemize}
\end{itemize}

\subsection{Interactive commands}

Interactive commands are invoked by the developer as code actions in the IDE, typically when the cursor is inside the definition of a feature (routine).
These commands may create intermediate files or artifacts in the workspace, but they do not modify the files currently open in the developer’s editor.
This approach minimizes disruption and reduces context switching.

\subsubsection{Add model-based contracts}\label{add-model-based-contracts}
This command tries to add model-based contracts to a given feature.
As its focus in only on specifications, it may only extend the feature's precondition and postcondition blocks.
It prompts an LLM following this schema, where the ``*'' before a placeholder means that it is kept as a placeholder in the final prompt.

\begin{lstlisting}
-- <TASK>
-- <MODEL CURRENT CLASS>
-- <MODEL ARGUMENT CLASSES>
-- <PRECONDITION AVAILABLE IDENTIFIERS>
-- <POSTCONDITION AVAILABLE IDENTIFIERS>
<FEATURE NAME> <PARAMETERS> <RETURN TYPE>
    <HEADER COMMENT>
    require
        <PREEXISTING PRECONDITIONS>
        *<ADD NEW PRECONDITIONS>
    do
        <ALREADY PRESENT BODY>
    ensure
        <PREEXISTING POSTCONDITIONS>
        *<ADD NEW POSTCONDITIONS>
    end
\end{lstlisting}

The user can include additional hints in the feature's header comment, which are incorporated into the prompt.
The LLM returns a set of candidate feature definitions.
Eiffel-tools parses each candidate to extract the suggested contracts and then applies filtering to remove any unsuitable contracts.
Specifically, we remove contracts which:

\begin{itemize}
    \item Are redundant
    \item Are not model-based, and thus unusable by AutoProof
    \item Contain invalid top-level identifiers
    \item Contain calls with the wrong number of arguments
\end{itemize}

Finally, Eiffel-tools instructs the LSP client to insert the remaining valid contracts into the feature, appending them after any existing contracts.

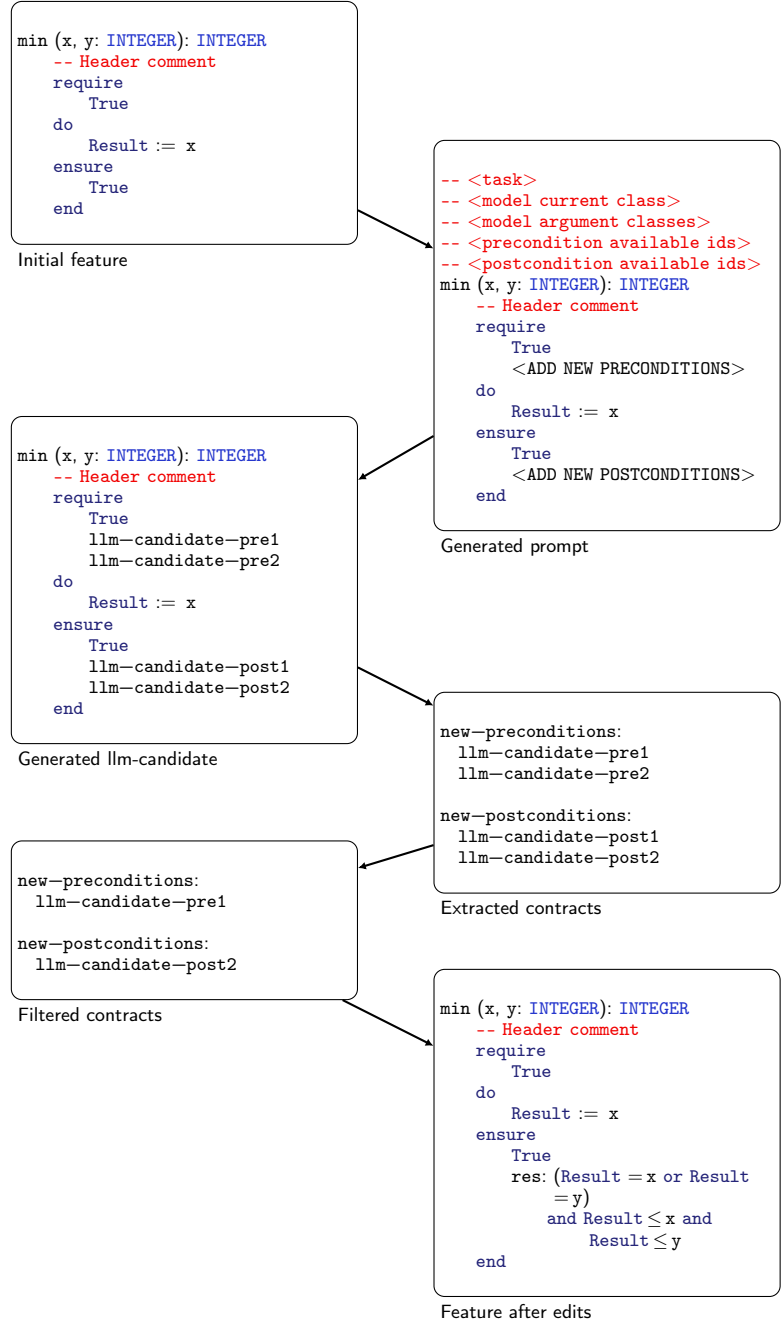
\begin{figure}
    \begin{tikzpicture}[every node/.style={fill=white, font=\sffamily, scale=0.8, text width=5.5cm}]
        \node (start) [code, label=below:Initial feature] {
            \begin{lstlisting}
min (x, y: INTEGER): INTEGER
    -- Header comment
    require
        True
    do
        Result := x
    ensure
        True
    end
\end{lstlisting}
        };

        \node (prompt) [code, right = of start, yshift=-100pt, label=below:Generated prompt] {
            \begin{lstlisting}
-- <task>
-- <model current class>
-- <model argument classes>
-- <precondition available ids>
-- <postcondition available ids>
min (x, y: INTEGER): INTEGER
    -- Header comment
    require
        True
        <ADD NEW PRECONDITIONS>
    do
        Result := x
    ensure
        True
        <ADD NEW POSTCONDITIONS>
    end
\end{lstlisting}
        };
        \draw[->, thick] (start) -- (prompt);

        \node (llm-candidate) [code, left = of prompt, yshift=-115pt, label=below:Generated llm-candidate] {
            \begin{lstlisting}
min (x, y: INTEGER): INTEGER
    -- Header comment
    require
        True
        llm-candidate-pre1
        llm-candidate-pre2
    do
        Result := x
    ensure
        True
        llm-candidate-post1
        llm-candidate-post2
    end
\end{lstlisting}
        };
        \draw[->, thick] (prompt) -- (llm-candidate);

        \node (extracted-contracts) [code, right = of llm-candidate,
        yshift=-100pt, label=below:Extracted contracts] {
            \begin{lstlisting}[language=]
new-preconditions:
  llm-candidate-pre1
  llm-candidate-pre2

new-postconditions:
  llm-candidate-post1
  llm-candidate-post2
\end{lstlisting}
        };
        \draw[->, thick] (llm-candidate) -- (extracted-contracts);

        \node (filtered-contracts) [code, left = of extracted-contracts, yshift=-60pt, label=below:Filtered contracts] {
            \begin{lstlisting}[language=]
new-preconditions:
  llm-candidate-pre1

new-postconditions:
  llm-candidate-post2
\end{lstlisting}
        };
        \draw[->, thick] (extracted-contracts) -- (filtered-contracts);

        \node (initial-feature-after-edits) [code, right = of filtered-contracts, yshift=-100pt,label=below:Feature after edits] {
            \begin{lstlisting}
min (x, y: INTEGER): INTEGER
    -- Header comment
    require
        True
    do
        Result := x
    ensure
        True
        res: (Result = x or Result = y)
            and Result <= x and Result <= y
    end
\end{lstlisting}
        };
        \draw[->, thick] (filtered-contracts) -- (initial-feature-after-edits);
    \end{tikzpicture}
    \caption{Dataflow of a feature in the add model-based contracts command}\label{dataflow-model-based-contracts}
\end{figure}

An example is shown at \figref{dataflow-model-based-contracts}.

\subsubsection{Fix feature body}\label{fix-feature-body}
This command generates an implementation for a feature that satisfies its model-based specifications.
It modifies only the feature's body, never its specifications.
Because AutoProof can verify only complete, compilable programs, Eiffel-tools handles intermediate results by creating a subclass of the current class and overriding the target feature in it.
The overriding feature is initially given to AutoProof.
If verification fails, \toolname{} enters a loop: it prompts the LLM for a new candidate body, inserts the candidate into the overriding feature, and reruns AutoProof.
If a candidate body passes verification, Eiffel-tools copies that body into the original feature; otherwise, the loop continues until a candidate passes or a maximum number of attempts is reached.
Unless a candidate succeeds, no changes are applied to the original feature.

\begin{figure}
    \begin{adjustbox}{max width=\linewidth, max totalheight=.9\textheight, keepaspectratio}
    \begin{tikzpicture}[every node/.style={font=\sffamily, scale=0.8, text width=7cm}]
        \node (start) [code, label=below:Initial feature in class] {
            \begin{lstlisting}
class
    EXAMPLE
feature
    min (x, y: INTEGER): INTEGER
        -- Header comment
        do
            Result := x
        ensure
            res: (Result = x or Result = y)
                and Result <= x and Result <= y
        end
end
\end{lstlisting}
        };

        \node (prompt) [code, right=of start, yshift=-80pt, label=below:Generated prompt] {
            \begin{lstlisting}
-- <task>
-- <model current class>
-- <model argument classes>
-- <precondition available ids>
-- <postcondition available ids>
min (x, y: INTEGER): INTEGER
    -- Header comment
    do
        <ADD BODY>
    ensure
        res: (Result = x or Result = y)
            and Result <= x and Result <= y
    end
\end{lstlisting}
        };
        \draw[->, thick] (start) -- (prompt);

        \node (llm-candidate) [code, left=of prompt, yshift=-80pt, label=below:Generated llm-candidate] {
            \begin{lstlisting}
min (x, y: INTEGER): INTEGER
    -- Header comment
    do
        if x < y then
            Result := x
        else
            Result := y
        end
    ensure
        res: (Result = x or Result = y)
            and Result <= x and Result <= y
    end
\end{lstlisting}
        };
        \draw[->, thick] (prompt) -- (llm-candidate);

        \node (extracted-body) [code, right=of llm-candidate, yshift=-50pt, label=below:Extracted body] {
            \begin{lstlisting}
if x < y then
    Result := x
else
    Result := y
end
\end{lstlisting}
        };
        \draw[->, thick] (llm-candidate) -- (extracted-body);

        \node (mid-artifact) [code, left=of extracted-body,
        yshift=-140pt, label=below:Instrumented feature redefinition] {
            \begin{lstlisting}
class
    LLM_INSTRUMENTED_EXAMPLE
inherit
    EXAMPLE redefine min end
feature
    min (x, y: INTEGER): INTEGER
        do
            if x < y then
                Result := x
            else
                Result := y
            end
        ensure
            res: (Result = x or Result = y)
                and Result <= x and Result <= y
        end
end
\end{lstlisting}
        };
        \draw[->, thick] (extracted-body.west) -- (mid-artifact);

        \node (verification-result) [code, right=of mid-artifact, yshift=-80pt] {Verification output};
        \draw[->, thick] (mid-artifact) -- (verification-result);
        \draw[->, thick] (verification-result.east) -- node[right=2pt] {Fails} ++(2,0) |- (prompt);
        \node (initial-feature-after-edits) [code, left=of verification-result, yshift=-125pt, label=below:Feature after edits] {
            \begin{lstlisting}
min (x, y: INTEGER): INTEGER
    -- Header comment
    require
        True
    do
        if x < y then
            Result := x
        else
            Result := y
        end
    ensure
        True
        res: (Result = x or Result = y)
            and Result <= x and Result <= y
    end
\end{lstlisting}
        };
        \draw[->, thick] (verification-result) -- (initial-feature-after-edits)
            node[midway, right] {Succeds};
    \end{tikzpicture}
    \end{adjustbox}
    \caption{Dataflow of a feature in the fix feature body command}\label{dataflow-feature-body}
\end{figure}
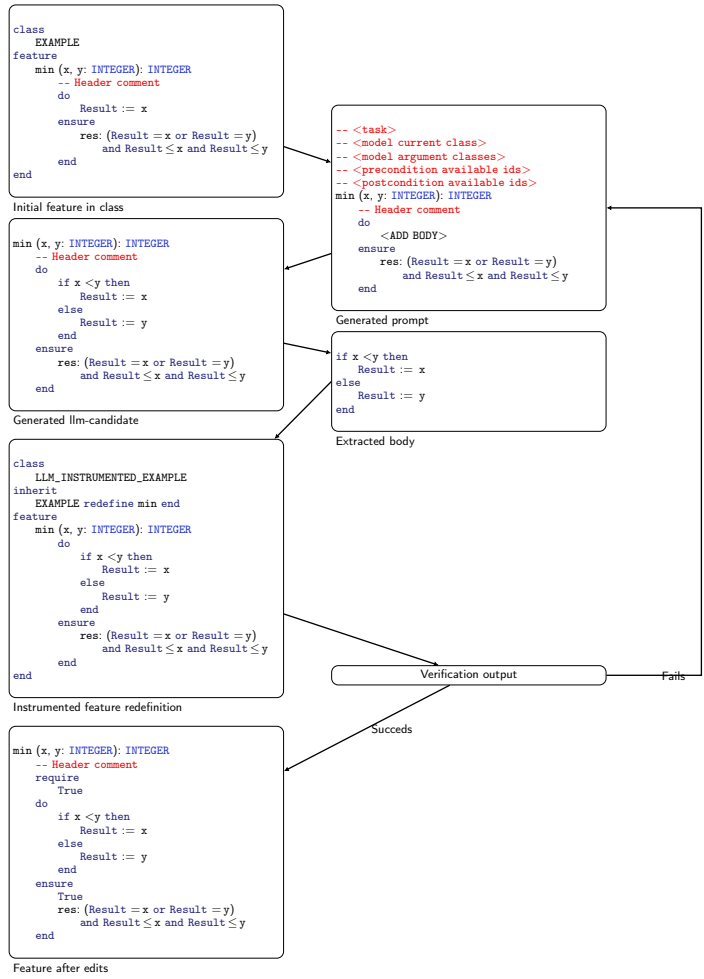

An example is shown at \figref{dataflow-feature-body}.

\subsection{Non-interactive commands}

Non-interactive commands run as standalone executables, enabling batch execution and clear tracking of LLM-generated changes under version control.
These commands may modify the project code and specifications directly, writing changes in place.
After each change, Eiffel-tools runs AutoProof: if AutoProof fails to verify the modified code, the tool reverts to the last stable state.

\subsubsection{Feature by feature body correction}\label{feature-by-feature}
This command takes an Eiffel configuration file and a list of classes or
features in classes.
It parses the entire project and attempts to verify each feature in the specified classes
or only the given features.
If a feature fails verification, the tool prompts the LLM to suggest a correction.
It may modify the code of the failing feature. The contracts cannot be changed
and assumed to be correct.
The LLM prompt includes the task description,
a list of available identifiers,
the feature's code,
the class invariant,
and the AutoProof counterexample (or a compilation error).

The prompt to the LLM consists of two messages. The first message is a \emph{system}
message including the task instructions and a short Eiffel syntax guide.
The second message is a \emph{user} message which includes the feature to be fixed
and the verification or compilation errors present in the current version of the code.

\begin{lstlisting}
System:
-- <TASK>

User:
-- <AVAILABLE IDENTIFIERS>
<FEATURE CODE>
-- <CLASS INVARIANT>
-- <AUTOPROOF COUNTEREXAMPLE>
\end{lstlisting}

From each LLM response, the parser extracts a candidate feature definition.
\toolname{} backs up the original feature, replaces it with the candidate, and reruns AutoProof.
If verification still fails, the original is restored and another candidate is tried.
This process repeats until a candidate verifies or a maximum number of attempts is reached.

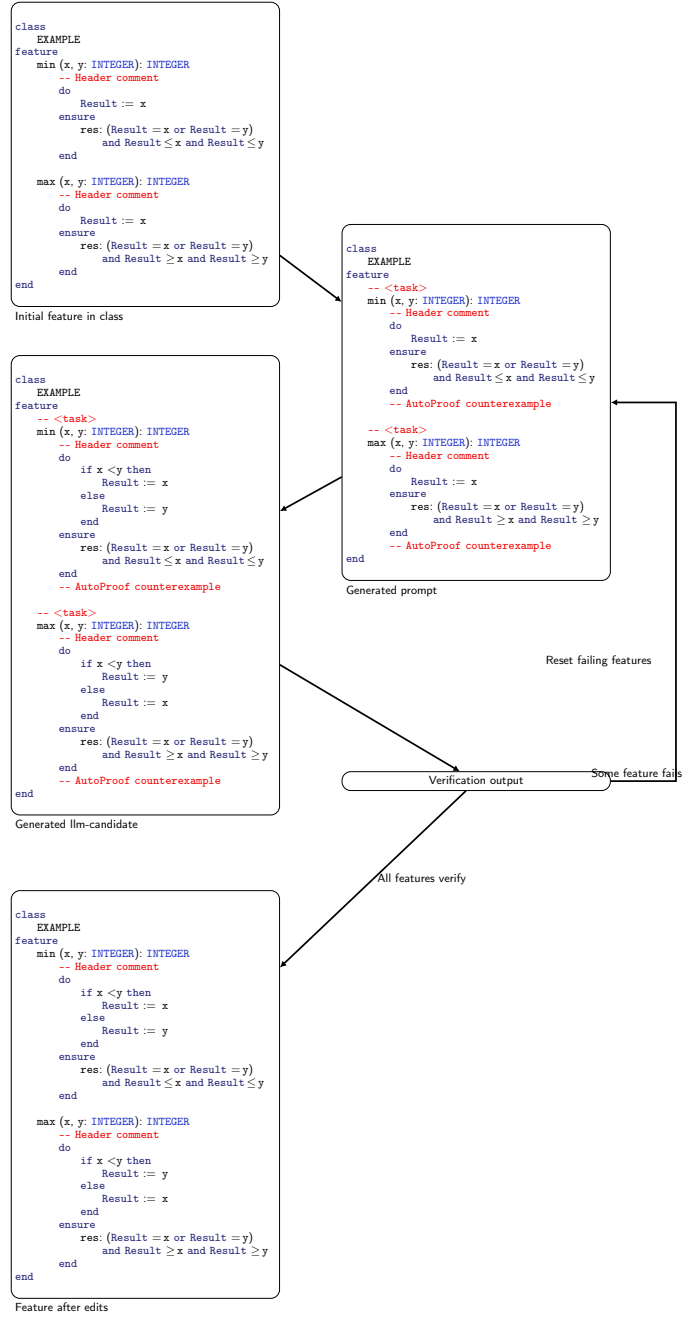
\begin{figure}
    \begin{adjustbox}{max width=\linewidth, max totalheight=.9\textheight, keepaspectratio}
    \begin{tikzpicture}[every node/.style={font=\sffamily, scale=0.6, text width=7cm}, align=left]
        \node (start) [code, label=below:Initial feature in class] {
            \begin{lstlisting}
class
    EXAMPLE
feature
    min (x, y: INTEGER): INTEGER
        -- Header comment
        do
            Result := x
        ensure
            res: (Result = x or Result = y)
                and Result <= x and Result <= y
        end

    max (x, y: INTEGER): INTEGER
        -- Header comment
        do
            Result := x
        ensure
            res: (Result = x or Result = y)
                and Result >= x and Result >= y
        end
end
\end{lstlisting}
        };

        \node (prompt) [code, right=of start, yshift=-190pt,label=below:Generated prompt] {
\begin{lstlisting}
class
    EXAMPLE
feature
    -- <task>
    min (x, y: INTEGER): INTEGER
        -- Header comment
        do
            Result := x
        ensure
            res: (Result = x or Result = y)
                and Result <= x and Result <= y
        end
        -- AutoProof counterexample

    -- <task>
    max (x, y: INTEGER): INTEGER
        -- Header comment
        do
            Result := x
        ensure
            res: (Result = x or Result = y)
                and Result >= x and Result >= y
        end
        -- AutoProof counterexample
end
\end{lstlisting}
        };
        \draw[->, thick] (start) -- (prompt);

        \node (llm-candidate) [code, left=of prompt, yshift=-140pt, label=below:Generated llm-candidate] {
            \begin{lstlisting}
class
    EXAMPLE
feature
    -- <task>
    min (x, y: INTEGER): INTEGER
        -- Header comment
        do
            if x < y then
                Result := x
            else
                Result := y
            end
        ensure
            res: (Result = x or Result = y)
                and Result <= x and Result <= y
        end
        -- AutoProof counterexample

    -- <task>
    max (x, y: INTEGER): INTEGER
        -- Header comment
        do
            if x < y then
                Result := y
            else
                Result := x
            end
        ensure
            res: (Result = x or Result = y)
                and Result >= x and Result >= y
        end
        -- AutoProof counterexample
end
\end{lstlisting}
        };
        \draw[->, thick] (prompt) -- (llm-candidate);

        \node (verification-result) [code, right=of llm-candidate, yshift=-150pt] { Verification output };
        \draw[->, thick] (llm-candidate) -- (verification-result);
        \draw[->, thick] (verification-result.east) -| ([xshift=30pt] prompt.east)
            node[midway, below, yshift=100pt] {Reset failing features}
            node[near start, above, xshift=60pt] {Some feature fails} -- (prompt.east);
        \node (initial-feature-after-edits) [code, left=of verification-result, yshift=-240pt, label=below:Feature after edits] {
            \begin{lstlisting}
class
    EXAMPLE
feature
    min (x, y: INTEGER): INTEGER
        -- Header comment
        do
            if x < y then
                Result := x
            else
                Result := y
            end
        ensure
            res: (Result = x or Result = y)
                and Result <= x and Result <= y
        end

    max (x, y: INTEGER): INTEGER
        -- Header comment
        do
            if x < y then
                Result := y
            else
                Result := x
            end
        ensure
            res: (Result = x or Result = y)
                and Result >= x and Result >= y
        end
end
\end{lstlisting}
        };
        \draw[->, thick] (verification-result) -- (initial-feature-after-edits)
            node[midway, right] {All features verify};
    \end{tikzpicture}
    \end{adjustbox}
    \caption{Dataflow of a feature in the feature by feature correction command}\label{dataflow-feature-by-feature-correction}
\end{figure}

An example is shown at \figref{dataflow-feature-by-feature-correction};

\subsubsection{Class by class correction}\label{class-by-class}
This command takes an Eiffel configuration file and a list of classes.
It verifies and repairs entire classes as a unit.
It may modify both the specifications and code of all features in a class.
After parsing the project, Eiffel-tools verifies each class as a whole.
If a class fails verification, the LLM is asked to generate a new candidate version of the entire class.
Eiffel-tools first backs up the original class.
If the LLM's candidate is syntactically correct, it replaces the class, and AutoProof is run on it.
If verification succeeds, Eiffel-tools moves on to the next class.
Otherwise, it reverts to the backup and retries with another candidate (up to a fixed limit).
This is the only command that attempts to produce a large, consistent output (a whole class) in one step.

Each feature in the class prompt is formatted with its associated counterexample, as follows:

\begin{lstlisting}
-- <TASK>
<FEATURE CODE>
-- <AUTOPROOF COUNTEREXAMPLE ABOUT FEATURE>
\end{lstlisting}

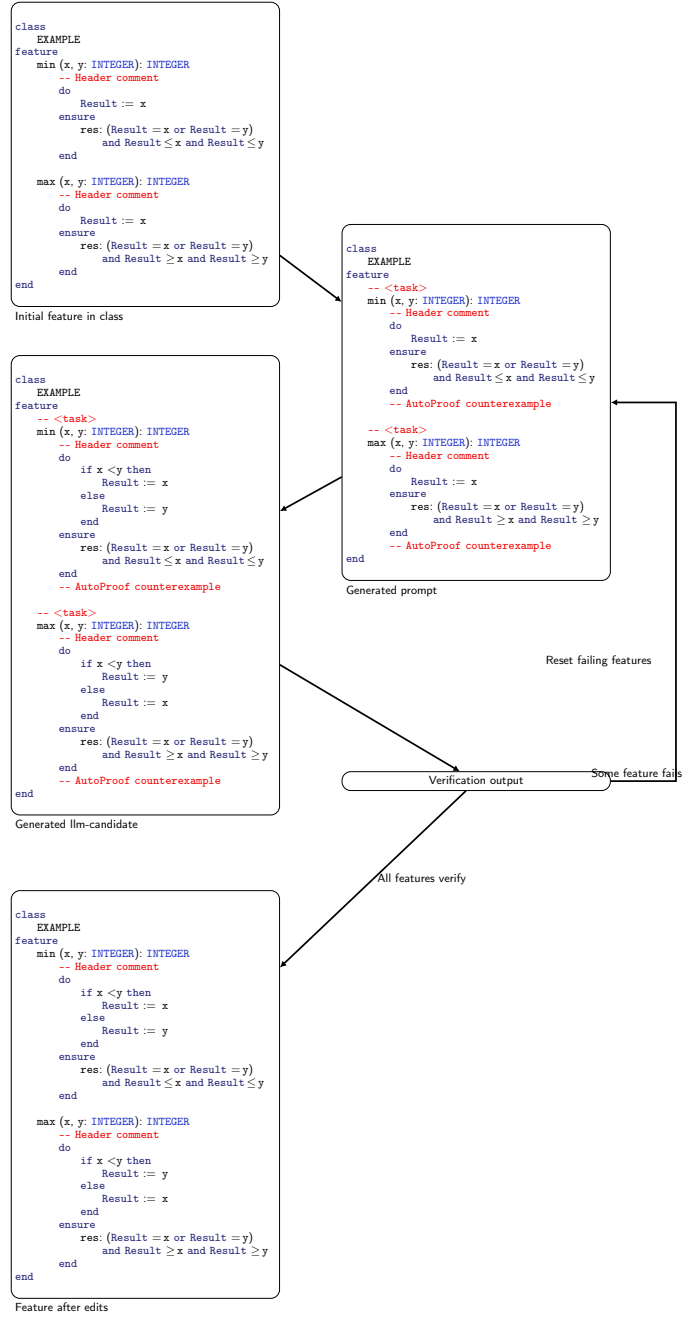
\begin{figure}
    \begin{adjustbox}{max width=\linewidth, max totalheight=.9\textheight, keepaspectratio}
    \begin{tikzpicture}[, every node/.style={font=\sffamily, scale=0.6, text width=7cm}, align=left]
        \node (start) [code, label=below:Initial feature in class] {
            \begin{lstlisting}
class
    EXAMPLE
feature
    min (x, y: INTEGER): INTEGER
        -- Header comment
        do
            Result := x
        ensure
            res: (Result = x or Result = y)
                and Result <= x and Result <= y
        end

    max (x, y: INTEGER): INTEGER
        -- Header comment
        do
            Result := x
        ensure
            res: (Result = x or Result = y)
                and Result >= x and Result >= y
        end
end
\end{lstlisting}
        };

        \node (prompt) [code, right=of start, yshift=-190pt, label=below:Generated prompt] {
\begin{lstlisting}
class
    EXAMPLE
feature
    -- <task>
    min (x, y: INTEGER): INTEGER
        -- Header comment
        do
            Result := x
        ensure
            res: (Result = x or Result = y)
                and Result <= x and Result <= y
        end
        -- AutoProof counterexample

    -- <task>
    max (x, y: INTEGER): INTEGER
        -- Header comment
        do
            Result := x
        ensure
            res: (Result = x or Result = y)
                and Result >= x and Result >= y
        end
        -- AutoProof counterexample
end
\end{lstlisting}
        };
        \draw[->, thick] (start) -- (prompt);

        \node (llm-candidate) [code, left=of prompt, yshift=-140pt, label=below:Generated llm-candidate] {
            \begin{lstlisting}
class
    EXAMPLE
feature
    -- <task>
    min (x, y: INTEGER): INTEGER
        -- Header comment
        do
            if x < y then
                Result := x
            else
                Result := y
            end
        ensure
            res: (Result = x or Result = y)
                and Result <= x and Result <= y
        end
        -- AutoProof counterexample

    -- <task>
    max (x, y: INTEGER): INTEGER
        -- Header comment
        do
            if x < y then
                Result := y
            else
                Result := x
            end
        ensure
            res: (Result = x or Result = y)
                and Result >= x and Result >= y
        end
        -- AutoProof counterexample
end
\end{lstlisting}
        };
        \draw[->, thick] (prompt) -- (llm-candidate);

        \node (verification-result) [code, right=of llm-candidate, yshift=-150pt] { Verification output };
        \draw[->, thick] (llm-candidate) -- (verification-result);
        \draw[->, thick] (verification-result.east) -| ([xshift=30pt] prompt.east)
            node[midway, below, yshift=100pt] {Reset failing features}
            node[near start, above, xshift=60pt] {Some feature fails} -- (prompt.east);
        \node (initial-feature-after-edits) [code, left=of verification-result, yshift=-240pt, label=below:Feature after edits] {
            \begin{lstlisting}
class
    EXAMPLE
feature
    min (x, y: INTEGER): INTEGER
        -- Header comment
        do
            if x < y then
                Result := x
            else
                Result := y
            end
        ensure
            res: (Result = x or Result = y)
                and Result <= x and Result <= y
        end

    max (x, y: INTEGER): INTEGER
        -- Header comment
        do
            if x < y then
                Result := y
            else
                Result := x
            end
        ensure
            res: (Result = x or Result = y)
                and Result >= x and Result >= y
        end
end
\end{lstlisting}
        };
        \draw[->, thick] (verification-result) -- (initial-feature-after-edits)
            node[midway, right] {All features verify};
    \end{tikzpicture}
    \end{adjustbox}
    \caption{Dataflow of the class by class correction command}\label{dataflow-class-by-class-correction}
\end{figure}

\section{Evaluation}\label{evaluation}
We evaluated the feature by feature body correction command,
described in \secref{feature-by-feature},
on the datasets buggy-java-jml-eiffel and maple-recursive-eiffel
using three LLMs, GPT-5 Nano, Claude Sonnet 4.6 and Poolside Laguna XS
with 2 ablations (versions) of prompts used.

For every bug fixing task, 10 attempts were made by \toolname{}
to fix the error. Each attempt included the initial verification,
prompting the model with the verification result, verifying the
suggested fix. If the suggested fix does not verify, it is kept
and the next attempt starts with the previously suggeste fix, enabling
a feedback loop between LLM and the verifier.

All evaluations are in \toolname{} repository within the corresponding
folder \texttt{experiment-results} \cite{eiffel-tools}.

\subsection{Models}
The experiment compares performance of several models in the bug fixing task.
All model were accessed through the OpenRouter API.

The Claude Sonnet 4.6 model from Anthropic \cite{ClaudeSonnet4.6} was created
especially for coding and for for building complex agents. This model
represents the ``expensive but powerful'' option in the experiment.

The GPT-5 Nano model from OpenAI \cite{openai-gpt5nano} is a generalist model
``great for summarization and classification tasks''. Comparing with Sonnet 4.6,
the nano model provides lower latency and a cheaper cost, so it represents the
``cheap but fast'' option in the experiment.

The Poolside Laguna XS~\cite{abadji2026laguna} model is a smaller model with
open weights. At the moment of writing, it was provided for free, so it represents
the ``free and open-weights'' option in the experiment.

\subsection{Datasets}
Performance of the bug fixing capability was measured on two publicly available
datasets of buggy programs with contracts (formal specifications).

The buggy-java-jml-eiffel dataset is a translation of the BuggyJava+JML dataset \cite{buggy-java-jml}
from Java and JML \cite{jml} to Eiffel.
The Eiffel translation is available online at \cite{Mustafin_buggy-java-jml-eiffel_2024}.
Examples in the dataset include
simple algorithms (function to return the absolute value of an integer),
array algorithms (copying an array, binary search),
data structures (stack, queue).
In total, 513 examples were used from the dataset.

The maple-recursive-eiffel dataset is a translation of the Maple dataset \cite{nguyen2019automatic} from C to Eiffel.
The translation was used as a part of the evaluation benchmark in Proof2fix \cite{proof2fix}
and is available online at \cite{Mustafin_maple-recursive-eiffel_2024}.
Examples in the dataset include
simple algorithms (absolute value, sum, min, max)
and recursive functions (recursive counter).
In total the dataset contains 26 examples and all of them were used in the
evaluation.

\subsection{Ablations}
To measure the usefulness of \toolname{} in helping LLMs work with a low-resource
language like Eiffel, performance of models is compared using
two prompt ablations (versions).

The ``full'' version includes all parts of the prompt described in \secref{feature-by-feature}.
System message explains the Eiffel syntax which makes the most problems to LLMs,
including the order clauses in the loop (\lstinline|from invariant until loop variant end|),
arithmetic operators (\lstinline|// \\| for div and mod respectively).
Additionally, system message provides the prefilled template (\figref{full-template}) for the expected
fixed feature.

User message in the ``full'' ablation lists the names and types of variables
available in pre- and post-conditions and includes the feature to be fixed and
the compilation or verification error from AutoProof.

\begin{figure}
\begin{lstlisting}[language={}, literate={...}{{...}}3]
Your response must be a single ```eiffel code block. The block must begin verbatim with the feature header (signature, documentation comments, notes, and precondition - exactly as shown). Then add your corrected local declarations and body. End the block with the verbatim postcondition and closing `end` before the closing ```.
Required structure:
```eiffel
div (a_n, a_d: INTEGER): INTEGER
		note
			status: functional
		require
			non_zero_divisor: a_d /= 0
... (your corrected local declarations and body here) ...
		ensure
			res_mod: Result = a_n \\ a_d
		end
```
\end{lstlisting}
\caption{Part of the ``full'' ablation system message specifying the feature template}\label{full-template}
\end{figure}

The ``minimal'' ablation omits almost all of the rich features extracted by the tool.
System message does not have the Eiffel syntax guide nor the verbatim feature template.
Instead, a natural text explanation is provided: ``The first line of your
response must be \texttt{```eiffel} and the second line must be the feature signature''.

User message omits the list of available variables and the error itself.
Model is only told in the system message that there is a problem,
but the exact problem is not specified in the user message.

\section{Results and Discussion}\label{results-and-discussion}
In total, the experiment used 539 bugs to be fixed by 3 LLMs
using 2 prompt ablations. At most 10 attempts were allowed per feature.

\begin{figure}
  \includegraphics{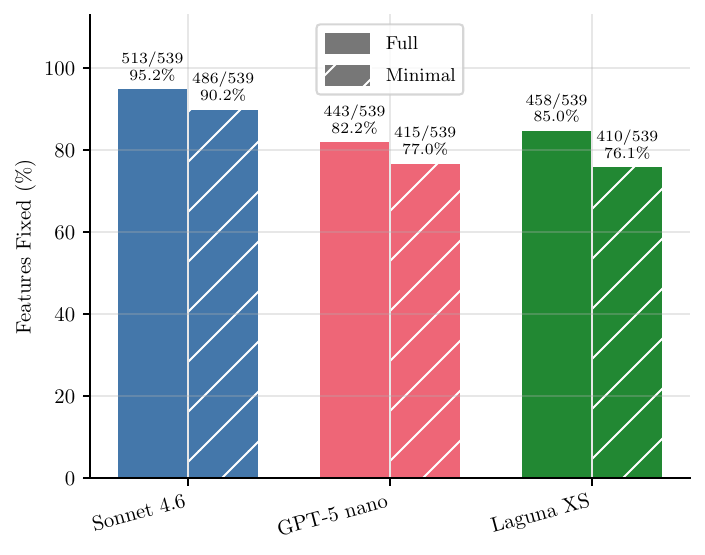}
  \caption{Model success rate}
  \label{fig-success-rate}
\end{figure}

\figref{fig-success-rate} shows how many bugs were fixed by each model in
every ablation.

Sonnet 4.6 solved the most tasks in both ablations (95\% full, 90\% minimal).
Laguna XS was better than GPT-5 nano in ``full'' (85\% vs. 82\%) while
GPT was better than Lagune in ``minimal'' (77\% GPT vs. 76\% Laguna).

Using ``full'' gives 6.4 pp improvement on average. Extended prompts helped
Laguna to improve the most (8.9 pp).

\begin{figure}
  \includegraphics{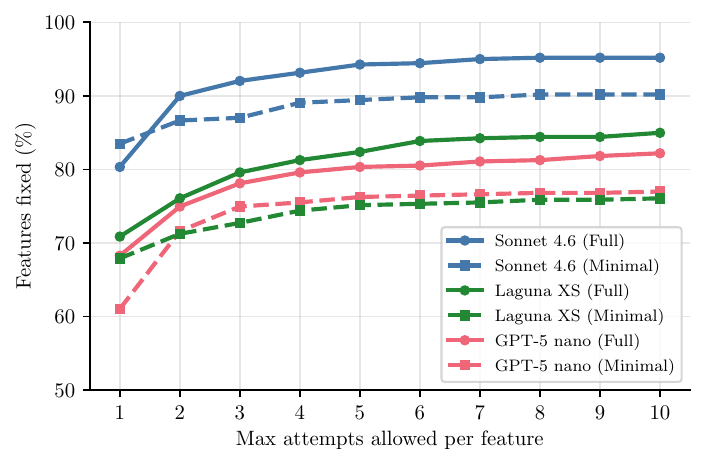}
  \caption{Model success rate per attempts (only fixed)}
  \label{fig-cumulative}
\end{figure}

\figref{fig-cumulative} shows the cumulative success rate per the number
of attempts used. The majority of fixes happen on the first attempt
(from 67.9\% for Laguna ``minimal'' to 83.5\% for Sonnet ``minimal'').
Every next attempt increases the success rate less and less.

The ``full'' ablation results in a higher success rate consistently starting
from the 2nd attempt.

\begin{figure}
  \includegraphics{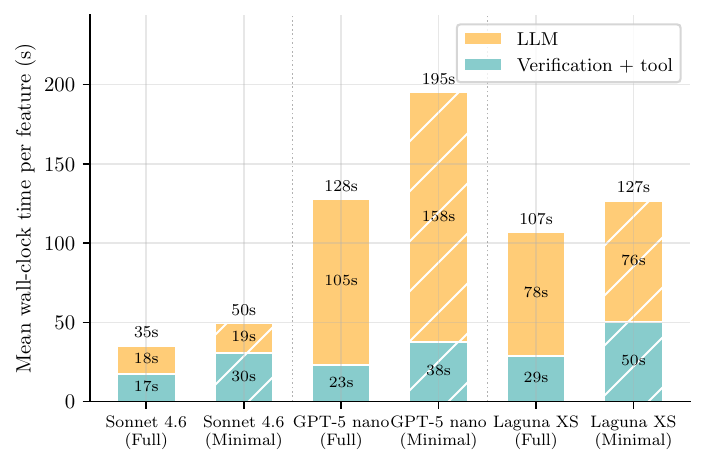}
  \caption{Mean time per feature (fixed and unfixed)}
  \label{fig-time}
\end{figure}

\figref{fig-time} shows the mean time used per feature run,
including both sucessfully fixed and still unfixed features
after all attempts are exhausted. The diagram shows the time spent
in the LLM (waiting for the model response) separately from the time
spent in other steps (verification, tool overhead).

LLM time depends on the model itself and the provider's API. While there
are several configurable settings in OpenRouter which can optimize throughput
and latency, the experiment used the default settings.

Experiments using smaller models spent the majority of time waiting for LLM
while Sonnet's LLM time was similar to verification + tool time.

Using the ``minimal'' ablation increases the total time per feature which
is mainly connected with the increase in the number of attempts used.

\begin{figure}
  \includegraphics{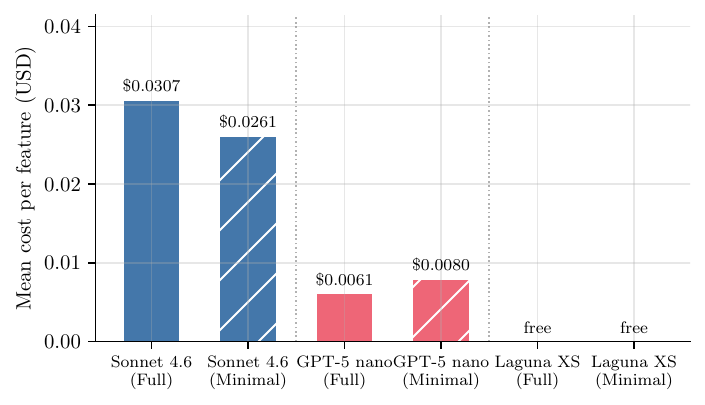}
  \caption{Mean cost per feature (fixed and unfixed)}
  \label{fig-cost}
\end{figure}

\figref{fig-cost} present the mean cost used per feature run,
including both sucessfully fixed and unfixed features.

Because Laguna was available for free at the time of writing and no pricing
information was known, only Sonnet 4.6 and GPT-5 nano are compared.

While Sonnet's usage per feature is more expensive than of GPT
(5x in ``full'' and 3.3x in ``minimal''),
Sonnet is faster (2x in ``full'', 3.3x in ``minimal'').

\begin{figure}
  \includegraphics{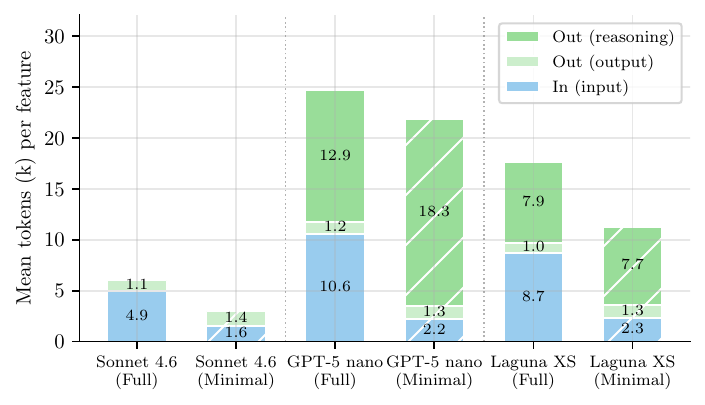}
  \caption{Mean input, output and reasoning tokens per feature (fixed and unfixed)}
  \label{fig-tokens}
\end{figure}

Switching to the ``minimal'' ablation decreased the cost per feature when
using Sonnet but increased when using GPT. The difference can be attributed
to the increase in attempts needed and
the increase in reasoning tokens (\figref{fig-tokens}).

In the experiment Sonnet had reasoning disabled (default settings) but GPT and Laguna
have reasoning enabled. Reasoning tokens have the same cost as output tokens and
cost more (5x more expensive for Sonnet, 8x more expensive for GPT).
Minimal ablation triggers additional reasoning in GPT which increases the cost.

\section{Limitations and threats to validity}\label{limitations}
The benchmarks used in our evaluation, described in~\secref{evaluation},
are public and available in multiple programming languages, either C and Eiffel or Java-JML and Eiffel.
It is possible that those benchmarks leaked in the training data of some LLMs and
the models could reproduce the solution they were trained on instead of ``fixing''
the bug.

\section{Conclusion and Future work}\label{future-work}
This paper presented an approach for using an LSP server as both the
LLM harness and the integration interface between the editor, verifier
and LLM at the same time. Implementation of several advanced code editing actions
combining LLMs and vormal verification is presented.

The experiment results discuss trade-offs and
set a baseline for the possible further improvements
across several axes: fix success rate, time, and cost.

Further analysis of prompt ablations is needed to identify optimal
initial prompt and the possible extension of the system
with ``tool calls'' and presenting an MCP (Model Context Protocol) interfaces
for the compilar and verifier.

Authors plan extending the implementation of \toolname{} and analysing
performance of new code actions.

\bibliographystyle{IEEEtran}
\bibliography{Bibliography}

\end{document}